\documentclass[letterpaper, 11pt]{article}
\usepackage[margin=1in]{geometry}
\usepackage{array}
\newcolumntype{P}[1]{>{\centering\arraybackslash}p{#1}}
\usepackage{url}
\usepackage{color,soul}
\usepackage{tablefootnote}
\usepackage[official]{eurosym}
\usepackage{subfig}
\usepackage{lipsum}
\usepackage{booktabs,caption}
\usepackage[flushleft]{threeparttable}
\usepackage{graphicx}
\usepackage{dcolumn}
\usepackage{siunitx}
\usepackage{setspace}
\usepackage{titling} % used to repeat the title on the abstract page
\usepackage{amsmath}
\usepackage{graphics}
\usepackage{tabularx}
\usepackage{multirow}
\usepackage{bigdelim}

\usepackage[semicolon]{natbib}
 \bibpunct[, ]{(}{)}{;}{a}{}{,}%
 \def\bibfont{\small}%
\newcommand\cites[1]{\citeauthor{#1}'s\ (\citeyear{#1})}
 
\begin{document}
%%%%%%%%%%%%%%%%

%\setcounter{page}{1}
%\DOI{10.1287/xxxx.0000.0000}%


%\RUNTITLE{}

\title{\textbf{How Much Ad Viewability is Enough? The Effect of Display Ad Viewability on Advertising Effectiveness\thanks{The authors acknowledge helpful comments from Elea Feit, PK Kannan, Carl Mela, and Bernd Skiera, as well as conference participants at EMAC 2017 and 2018 and EMAC 2018 Doctoral Colloquium, Marketing Science Conference 2017 and 2018, Passau Digital Marketing Conference 2018, Theory + Practice in Marketing (TPM) Conference 2018, and seminar participants at Goethe University Frankfurt 2018 and Facebook 2018. The authors did not receive any third-party funding for this project. All remaining errors are our own.}}}

%\ARTICLEAUTHORS{%
% \author{Christina Uhl}
% \AFF{Vienna University of Economics and Business (WU), \EMAIL{christina.uhl@wu.ac.at}, \URL{}}
% \author{Klaus Miller}
% \AFF{Vienna University of Economics and Business (WU), \EMAIL{klaus.miller@wiwi.uni-frankfurt.de}, \URL{}}
% \AUTHOR{Nadia Abou Nabout}
% \AFF{Vienna University of Economics and Business (WU), \EMAIL{nadia.abounabout@wu.ac.at}, \URL{}}
% % Enter all authors
% } % end of the block

\author{
\textbf{Christina Uhl}\\
WU Vienna\\
\texttt{christina.uhl@wu.ac.at}\\\\\
\textbf{Nadia Abou Nabout}\\
WU Vienna\\
\texttt{nadia.abounabout@wu.ac.at} \\\\\
\textbf{Klaus M. Miller}\\
Goethe University Frankfurt\\
\texttt{klaus.miller@wiwi.uni-frankfurt.de}\\\\\
}

\pagenumbering{gobble}
\maketitle 

\newpage
\pagenumbering{arabic}

\begin{center}
\Large
\thetitle
\end{center}

\vspace{1in}

\begin{abstract}
A large share of all online display advertisements (ads) are never seen by a human. For instance, an ad could appear below the page fold, where a user never scrolls. Yet, an ad is essentially ineffective if it is not at least somewhat viewable. Ad viewability - which refers to the pixel percentage-in-view and the exposure duration of an online display ad - has recently garnered great interest among digital advertisers and publishers. However, we know very little about the impact of ad viewability on advertising effectiveness. We work to close this gap by analyzing a large-scale observational data set with more than 350,000 ad impressions similar to the data sets that are typically available to digital advertisers and publishers. This analysis reveals that longer exposure durations ($>$10 seconds) and 100\% visible pixels do not appear to be optimal in generating view-throughs. The highest view-through rates seem to be generated with relatively lower pixel/second-combinations of 50\%/1, 50\%/5, 75\%/1, and 75\%/5. However, this analysis does not account for user behavior that may be correlated with or even drive ad viewability and may therefore result in endogeneity issues. Consequently, we manipulated ad viewability in a randomized online experiment for a major European news website, finding the highest ad recognition rates among relatively higher pixel/second-combinations of 75\%/10, 100\%/5 and 100\%/10. Everything below 75\% or 5 seconds performs worse. Yet, we find that it may be sufficient to have either a long exposure duration \textit{or} high pixel percentage-in-view to reach high advertising effectiveness. Our results provide guidance to advertisers enabling them to establish target viewability rates more appropriately and to publishers who wish to differentiate their viewability products.

\end{abstract} 
%
%\newpage

%\tableofcontents

\newpage
\doublespacing
%%%%%%%%%%%%%%%%%%%%%%%%%%%%%%%%%%%%%%%%%%%%%%%%%%%%%%%%%%%%%%%%%%%%%%
% Samples of sectioning (and labeling) in MKSC
% NOTE: (1) \section and \subsection do NOT end with a period
%       (2) \subsubsection and lower need end punctuation
%       (3) capitalization is as shown (title style).
%
%\section{Introduction.}\label{intro} %%1.
%\subsection{Duality and the Classical EOQ Problem.}\label{class-EOQ} %% 1.1.
%\subsection{Outline.}\label{outline1} %% 1.2.
%\subsubsection{Cyclic Schedules for the General Deterministic SMDP.}
%  \label{cyclic-schedules} %% 1.2.1
%\section{Problem Description.}\label{problemdescription} %% 2.
% Text of your paper here

\section*{Introduction}\label{sec:intro}
According to the Interactive Advertising Bureau \citep{iab2019internet}, online advertising spending in the United States totaled \$124.6 billion in 2019, an increase of 15.9\% compared with 2018. Consequently, both researchers and practitioners have become greatly interested in online display advertising that enables user-level targeting and tracking \citep{choi2019online, eckles2018field, goldfarb2011online, goldfarb2014standardization, gordon2019comparison, johnson2016less, lambrecht2013does}. Most online advertising consists of display advertising in several formats, such as banner advertisements (ads), videos, and so forth. However, these formats invoke serious concerns over ad viewability; a measure of the extent to which display ads are viewable to online users \citep{emarketer2016a}. The common industry practice is to pay for ads served, regardless of where they appear on the page, according to a cost-per-mille (CPM).\footnote{Other well-known pricing methods for online display advertising are cost-per-click or cost-per-action, which are mainly used for performance-based advertising campaigns with a strong sales focus.} Yet, an ad served does not necessarily mean that it has been viewed; the ad could appear below the page fold, where a user never scrolls. Case in point: Industry reports suggest that more than half of all display ads are never seen by a human \citep{Bounie2017, marvin2014, vollman2013}.

However, relative to traditional advertising, display ad viewability has the advantage of being measurable: Viewability tracking software precisely documents how many pixels of each ad is within the user's view and for how many seconds. This technology makes the display advertising market more transparent, and accordingly, advertisers who care about display ad viewability demand that publishers document the viewability of advertising displayed on their websites, rather than just served impressions. As a result, new trading currency relies on viewable CPM (vCPM), with pricing established by the number of impressions viewed by Internet users. These attention-based pricing models for display banner ads have gained even greater importance since Google announced that it would restrict third-party tracking in its Chrome browser \citep{joseph2020}. Advertisers are increasingly concerned that the tracking of advertising effectiveness metrics (e.g., clicks, conversions) will be limited by the restriction of third-party cookies. Against this backdrop, advertisers have increasingly turned to the metric of viewability.\footnote{Note that restrictions or a complete ban of third-party tracking (e.g., based on cookies) will also affect third-party ad viewability measurement providers, which attempt to measure ad viewability across sites. However, such restrictions will not affect first-party ad viewability measurement providers, who track an online user only on the focal publisher's website.}  

Yet, the big question for advertisers is: How much ad viewability is enough from an advertising effectiveness stance? Targeting a maximum ad viewability may be tempting, but expensive as publishers may charge advertisers depending on ad viewability. In addition, ads might be effective even if they are not fully viewable. Specifically, one might believe that more viewability is always better, which might be justified by a mere exposure effect and perceptual fluency \citep{fang2007examination}. On the other hand, fully viewable ads might come across as obtrusive leading to reactance behavior \citep{goldfarb2011online}. Finally, one might even find some sort of saturation plausible \citep{vriens2006handbook}, where ad viewability increases advertising effectiveness until a certain threshold is reached. Because these scenarios would lead to vastly different recommendations for advertisers and publishers, we need a better understanding of the effects of ad viewability on advertising effectiveness. On this point, the extant marketing literature has uncovered some effects of exposure duration on advertising effectiveness, particularly in relation to ad recall for print ads \citet{houston1987picture} and ad recall and evaluations for online ads \citep{goldstein2011effects, pieters2012ad, elsen2016thin}. However, the precise impact of the number of pixels-in-view, in combination with exposure duration, has not been analyzed. For this reason, \citet{kannan2016path} and \citet{choi2019online} cited the effect of ad viewability metrics as an imminent question for marketing researchers. We work to close this gap by answering how the aforementioned combination of pixel percentage-in-view and exposure duration affects display advertising effectiveness. Because ad viewability is a prerequisite for attention, this topic is also of great interest to research on the so-called ``attention economy'' \citep{orquin2020contributions}. At the extreme, there is no attention without ad viewability.  

To answer our research question, we analyzed a large-scale observational data set that is typically accessible to digital advertisers and publishers and features more than 350,000 ad impressions (Study 1). This analysis revealed that longer exposure durations ($>$10 seconds) and 100\% of the pixels-in-view do not seem to be optimal in generating view-throughs (defined as a visit to the advertiser's website on the same day as the ad exposure). The highest view-through rates are generated with relatively lower pixel/second-combinations of 50\%/1, 50\%/5, 75\%/1, and 75\%/5. However, such an analysis can only weakly account for potential endogeneity concerns that stem from the fact that user behavior and specific surfing goals may be correlated with or even drive ad viewability. Consequently, we conducted an online experiment (Study 2) under controlled conditions and exogenously manipulated ad viewability. Participants responded to a post-exposure survey about ad recognition. Here, we see that relatively higher pixel/second-combinations of 75\%/10, 100\%/5 and 100\%/10 showed the highest ad recognition rates. Everything below 75\% or 5 seconds seems to perform worse. Yet, we find that it may be sufficient to have either a long exposure duration \textit{or} high pixel percentage-in-view to reach high advertising effectiveness. 

\section*{Definition and Recent Market Developments}\label{sec:definition}
Using modern advertising technologies, publishers and advertisers increasingly rely on ad servers to provide automatic ad delivery. To determine how many ads get delivered to a website's users, the systems use ad tags--small pieces of HTML or JavaScript code placed on each ad. The ad tags often come from an ad viewability measurement provider that tracks the number of ad impressions served and viewed by users of a specific website. The number of served impressions is the number of tagged impressions. However, not all served impressions get measured by these ad viewability measurement providers, due to network failures and invalid (non-human, bot) traffic issues. Thus, a second measure of the number of impressions is used, to clean the data of invalid traffic and non-served impressions. Even in this case, though, correctly served ads might not reach users, because they appear below the fold (i.e., outside the viewable browser space) or are blocked by an ad blocking software installed in the user's browser. Consequently, according to the Media Rating Council's (MRC) Viewable Ad Impression Measurement Guidelines, which initially distinguished a viewable ad impression from an impression served \citep{mrc2014, mrc2016}, ``a served ad impression can be classified as a viewable impression if the ad is contained in the viewable space of the browser window, on an in focus browser tab, based on pre-established criteria such as the percent of ad pixels within the viewable space and the length of time the ad is in the viewable space of the browser'' \citep{mrc2014}. The viewability rate refers to the ratio of the number of viewable impressions over the number of measured impressions.

The MRC suggests that display banner ads can be considered viewable if 50\% of their pixels are in view for a minimum of 1 second; for desktop video, it must be 50\% of pixels-in-view for 2 seconds. In addition, for larger desktop ads (i.e., 242,500 pixels, equivalent to the size of a 970 $\times$ 50 pixel display ad, or larger), 30\% of pixels must be in view for 1 continuous second \citep{mrc2014}. For mobile ads, the MRC recommends treating smartphone and desktop ads similarly: 50\% of the pixels must be in view for at least 1 continuous second. The MRC argues that a viewable ad impression is one that \textit{could} be seen by Internet users, regardless of whether it actually is seen.

While these standards represent guidelines and not regulations, several major industry players (e.g., comScore, and Google) have adopted them. For example, in real-time bidding auctions that allow advertisers to purchase ad inventory, marketers can target a specific viewability rate across publishers and ad placements. This viewability rate reflects the percentage of ad impressions counted as viewable, according to the MRC 50\%/1 second standard (which we abbreviate hereafter as 50/1), while targeting relies on predictions based on prior empirical data \citep{doubleclick2018}. Some platforms, such as Google, allow advertisers to pay solely for viewable impressions (according to the MRC standard), which gives advertisers a choice of two pricing schemes: a traditional CPM method or a method that charges solely for viewable impressions, or vCPM \citep{GoogleAdWords2018}. Google thereby recommends advertisers to bid higher when placing a bid for viewable impressions because those are more attractive to advertisers. In a cooperation of Meetrics, a leading ad verification provider, and Platform161, advertisers are even able to target expected viewability by combining three levels of viewable minimum areas (50\%, 75\%, or 90\% of pixels-in-view) with four minimum duration classes \citep[1, 3, 5, or 10 seconds;][]{Rowntree2016}. Furthermore, even as many companies adopt the MRC standard, some propose going beyond it: GroupM, for example, defines an ad as viewable only if 100\% of the pixels are in view for at least 1 second. \textit{The Financial Times} and \textit{The Economist} both offer a viewability-based pricing scheme, only charging for ads if they meet the 50\% pixel criterion and are in view for a minimum of 5 seconds \citep{Davies2015, Moses2015, Sanghvi2015}. Another ad management platform has suggested selling ad inventory using time-based pricing schemes according to a cost-per-second trading model \citep{Rowntree2017}.

Our findings clearly speak to these recent industry developments. While the highest view-through rates appear to be generated with relatively lower pixel/second-combinations of 50\%/1, 50\%/5, 75\%/1, and 75\%/5 in the observational study, the randomized online experiment reveals that relatively higher pixel/second-combinations of 75\%/10, 100\%/5 and 100\%/10 showed the highest ad recognition rates. Everything below 75\% or 5 seconds seems to perform worse. The observational study therefore suggests that the MRC standard is sufficient to ensure high advertising effectiveness. However, our randomized online experiment does not support this finding. Instead it shows that one either needs a long exposure duration (i.e., 10 seconds) \textit{or} high pixel percentage-in-view (i.e., 100\%) to reach high advertising effectiveness. Given that users spend on average about 2-3 seconds on a site \citep{Baker2017, digishuffle2019, albright2019}, these findings are important for advertisers and publishers alike. Advertisers may demand target viewability rates that go beyond the MRC standard. Publishers may decide to differentiate their viewability products further with the MRC standard just being their most basic product.

\section*{Related Literature and Contribution}\label{sec:lit_rev}
By investigating ad viewability in the context of online display advertising, the current research adds to three streams of literature.

First, our research addresses the general role of display ad exposures in online advertising effectiveness. For example, \citet{manchanda2006effect} studied the effect of display ad exposures on purchase behavior; the authors found that the number of ad exposures, websites visited, and individual webpages visited all have positive effects on repeat purchase probabilities, whereas the number of unique creatives has a negative effect. In \cites{rutz2012does} assessment of the effect of display ads on a user’s website browsing behavior, they found that display ads influence subsequent browsing choices during the current browsing session, but do not affect later browsing sessions. \citet{sahni2015effect} also outlined the impact of temporal spacing between ad exposures on purchase likelihood.

Second, one important dimension of an ad exposure is its duration, which is why we draw on research in this domain. In particular, we note \cites{goldstein2011effects} finding that exposure duration exerts a strong, causal influence on ad recognition and recall for durations of up to one minute, but diminishing marginal effects at longer durations. Similarly, \citet{pieters2012ad} investigated the impact of adverse exposure conditions (i.e., single eye fixation) on advertising effectiveness, demonstrating that consumers already know whether something is an ad or editorial material, as well as which product is being advertised, after an exposure of less than 100 milliseconds (ms). Even after an extremely coarse visual presentation of 100 ms, participants identify the advertised product and brand at well above chance levels. \citet{elsen2016thin} took this analysis further to show that upfront ads, which instantly convey what they promote, prompt more positive evaluations after both brief and longer exposure durations. Mystery ads, which avoid conveying what they promote, evoke negative views after brief exposures, but positive assessments after longer exposures. False front ads, which initially convey a different identity, reveal opposite trends: positive evaluations after brief exposures but negative ones after longer exposures.

Third, we investigate a novel dimension of ad exposures: pixels-in-view. Thus, we contribute to the relatively small, but growing literature on ad viewability. Despite this factor's practical importance \citep{emarketer2016a}, few academic studies challenge the assumption that ads are always viewable. One important sub-domain addresses how to predict ad viewability \citep{wang2015viewability,wang2017probabilistic,wang2018webpage,wang2017viewability}. Furthermore, \citet{Bounie2017} and \citet{bounie2016advertising} studied how adopting ad viewability technologies alters the economics of online advertising in a theoretical setting. Finally, only a few studies have sought to investigate the impact of ad viewability on online advertising effectiveness: Using eye-tracking devices, \citet{zhang2015empirical} determined the impact of ad viewability on ad recall. However, their study had only 20 participants and did not exogenously vary ad viewability across experimental participants. \citet{hill2015measuring} measured the effect of display ad exposures on advertising effectiveness, with ad viewability as a mediator, and revealed that approximately 45\% of display ad impressions never make it into a viewable portion of the user's browser. Including ad viewability as a mediator also substantially reduces the amount of bias in estimates of campaign lift. \citet{ghose2016towards} used a similar framework to empirically demonstrate that exposure to a display ad influences a wide range of consumer responses, including clicks, direct visits, search triggered visits, and conversions. Granted, it is difficult to identify causal effects because display ads are highly targeted \citep{lewis2015unfavorable}. Nonetheless, \citet{ghose2016towards} argued that actual ad viewability serves as an exogenous shock and simulates a quasi-experiment by creating two groups: users who do and do not view the ad. The authors argue that the assignment to these groups is random because both groups are initially targeted by the same marketing campaign. Thus, they can serve as treatment and control groups in an individual-level difference-in-differences analysis. These authors focus on the effect of exposure duration and found that longer exposures increase the likelihood that consumers directly visit the website rather than performing a search engine inquiry. However, their study did not consider the effect of pixels-in-view.

We seek to extend these different research streams in three main ways. First, previous studies focused on the effect of exposure duration on advertising effectiveness; we add the important dimension pixel percentage-in-view to this discussion. Second, to determine the causal effects of both these ad viewability dimensions on advertising effectiveness, we exogenously vary both ad viewability dimensions in an experimental setting. Third, this study is the first to compare observational and experimental data on ad viewability. With this effort, our results contribute to the academic literature on advertising effectiveness and ad viewability, but also provide advertisers and publishers with guidance on the ideal level of viewability.

\section*{Empirical Studies}\label{sec:emp_studies}
\subsection*{Measuring the Causal Effect of Ad Viewability on Advertising Effectiveness}
For our research, we relied on two unique data sets. In the observational study (Study 1), we analyzed view-throughs (i.e., a visit to the advertiser's website on the same day as the ad exposure) for various ad viewability values. View-throughs are a common industry metric and have already been analyzed by \citet{bleier2015personalized} in the context of individually personalized banners. 

This type of observational data is readily accessible to advertisers and publishers who work together with a viewability tracking software provider. However, using this observational data to quantify the causal effect of ad viewability on advertising effectiveness is a non-trivial task \citep[see][on the overall challenges of measuring advertising effectiveness]{gordon2019comparison, johnson2016less, eckles2018field}.

The reason is that ad viewability is the outcome of multiple factors: (1) user behavior, (2) website design and ad placement, (3) display size and browser settings, and (4) ad size and format. First, any given ad placement on a website might be more or less viewable depending on whether and how quickly the user scrolls down the page or leaves the website (e.g., by clicking on a link or closing the browser). In other words, an Internet user self-selects into different ad viewability classes through her browsing behavior. Second, an ad might be more or less viewable depending on where it is placed and how the website itself is designed. Some ad placements are located above the page fold and others below, which drives viewability. But even above-the-fold placements do not guarantee viewability, since website visitors might scroll down too quickly, making below-the-fold placements more viewable than those at the very top of the page \citep{Moses2013}. Third, display size and browser settings drive ad viewability. A user with a 17.3-inch laptop may be exposed to larger parts of a website and has a higher probability of seeing an ad compared to a user with a 11.6-inch laptop. In addition, individual browser settings may have an influence on ad viewability. A user who adopts a larger font size, is zoomed in on the website, or simply decreases the size of the browser window is exposed to smaller parts of the website, which then lowers the probability of an ad being in the visible area of the browser window. Finally, smaller ads might have a higher propensity of being viewable as they fit more easily into the browser's viewable area. 

Since ad viewability is driven by the factors outlined above, answering the question of how ad viewability affects advertising effectiveness is not straightforward. Indeed, one can easily reduce noise in the data by only looking at one specific ad placement on a certain website and under very specific browser settings. However, it is less clear how user behavior should be dealt with since it drives ad viewability and might additionally be correlated with advertising effectiveness. For instance, a user might go to a website to read a news article.  In that case, the user's time spent on the webpage - and by extension, ad viewability - will very likely be high. However, even though the ad was fully viewable, advertising effectiveness might have been low because that specific user had the goal of reading the news article and was not receptive to ads. Therefore, it might be misleading to attribute the low advertising effectiveness to ad viewability. Yet, advertisers are often limited to data that suffers from such shortcomings. The big question is whether analyzing observational data (Study 1) would result in substantially different recommendations than an actual, randomized experiment (Study 2). 

Given the shortcomings that come with observational data, conducting an exogenous variation of ad viewability (i.e., ensuring that user behavior cannot drive viewability) is important for understanding the causal effect of ad viewability on advertising effectiveness. To this end, we ran an online experiment (Study 2) in which we control for user behavior by exogenously manipulating exposure duration and pixel percentage-in-view. We then measured ad recognition in a post-exposure survey.\footnote{Since click-through and view-through rates for display ads are quite low (e.g., often fewer than 1 click per 1,000 ad impressions), these metrics would be very difficult, if not impossible, to study in an experimental setting with a limited number of study participants.} In order to further reduce noise in the data, we decided to study a very specific, but popular, setting in the online experiment: Specifically, we use the website of a nationwide online news website and their most commonly booked ad format since innovative viewability products have first been developed in the news industry (e.g., by the textit{The Financial Times} and \textit{The Economist}). 

Both studies include display ads from telecommunication providers: Study 1 uses data from the display ad campaign of a major European telecommunications provider that ran from January to April 2016. Study 2, which was conducted in November 2017, features an ad from a fictional brand that we specifically designed for this research.

\subsection*{Study 1: Observational Study}\label{sec:observational_data}
The type of data we employed in Study 1 is often readily available to advertisers and publishers who use viewability tracking software. To conduct the analysis, we had to combine display advertising data with website visit data (known as view-throughs). Our data comes from a major European telecommunications provider; the ad viewability of each individual ad exposure was tracked by a well-known ad viewability measurement provider. 

\subsubsection*{Data}
A European media planning agency collected the data between January and April 2016. This data collection produced 371,954 individual ad impressions by 17,480 unique users across 28 campaigns. The data contain 41 ad formats and includes above- and below-the-fold placements. The websites in our data set range from categories such as news and general interest to special interest as well as sites sold via programmatic advertising where the exact URL is unknown to the advertiser. To measure ad viewability, we obtained 10 viewability classes which contain distinct combinations of exposure duration and pixel percentage-in-view from the viewability tracking software provider, involving three minimum viewable areas (50\%, 75\%, and 100\% of pixels-in-view) and three minimum duration classes (1 second, 5 seconds, and 10 seconds of exposure duration). In the following, we will refer to the 10 resulting viewability classes as $<$50/1, 50/1, 50/5, 50/10, 75/1, 75/5, 75/10, 100/1, 100/5 and 100/10. Unlike in Study 2 (see below), we cannot confirm exact viewability with this data set. The actual viewability of these ads could lie anywhere between the respective classes for which we measure viewability; for example, an ad with a viewability of 50/5 could have an actual pixel percentage-in-view of anywhere between 50\% and 74\% and an actual exposure duration of anywhere between 5 and 9.99 seconds. An ad that doesn't reach a minimum of 50\% of pixels-in-view or a minimum of 1 second of exposure duration ends up in the $<$50/1 category. Even though not ideal, this type of measurement corresponds to common industry reporting standards.

\begin{table}[h]
\centering
\caption{Frequency Table for Viewability Values in Observational Data}
\label{table:n_obs}
\footnotesize
\begin{tabular}{p{2.5cm}lrr}
\toprule 
\textbf{Viewability} & \textbf{Frequencies} & \textbf{Shares} \\ 
\midrule
\textbf{$<$50\textbackslash 1}       & 113,153 & 30.4\%     \\
\textbf{50\textbackslash 1}        & 9,397   & 2.5\%     \\
\textbf{50\textbackslash 5}        & 5,615   & 1.5\%       \\
\textbf{50\textbackslash 10}       & 13,848  & 3.7\%       \\
\textbf{75\textbackslash 1}        & 32,918   & 8.9\%     \\
\textbf{75\textbackslash 5}        & 9,217   & 2.5\%       \\
\textbf{75\textbackslash 10}       & 28,782  & 7.7\%       \\
\textbf{100\textbackslash 1}        & 52,967   & 14.2\%     \\
\textbf{100\textbackslash 5}        & 23,910   & 6.4\%       \\
\textbf{100\textbackslash 10}       & 104,867  & 28.2\%       \\
\midrule
\textbf{Sum}       & 394,674  & 100\%       \\
\bottomrule
\end{tabular}

\begin{tablenotes}
      \footnotesize
      \item Notes: 371,952 observations. Overall number of viewability values does not equal overall number of ads because one ad exposure can exhibit multiple viewability values.
    \end{tablenotes}
\end{table}

Table \ref{table:n_obs} gives an overview of the viewability of our ads in the observational data set. Evidently, 30.4\% of the ads have a viewability value below 50/1 and are non-viewable by the MRC standard. A relatively big share of the ads, however, has a good viewability in excess of of 100/10 (i.e., 28.2\%). In general, the proportion of ads with 100\% of their pixels-in-view is relatively large in this specific data set (i.e., 48.8\% of total).

\subsubsection*{Model-free Evidence}
\begin{figure}[!htbp]
  \caption{View-through-rates per Viewability Value - Model-free Evidence from Observational Data}
  \centering
    \includegraphics[width=0.8\textwidth]{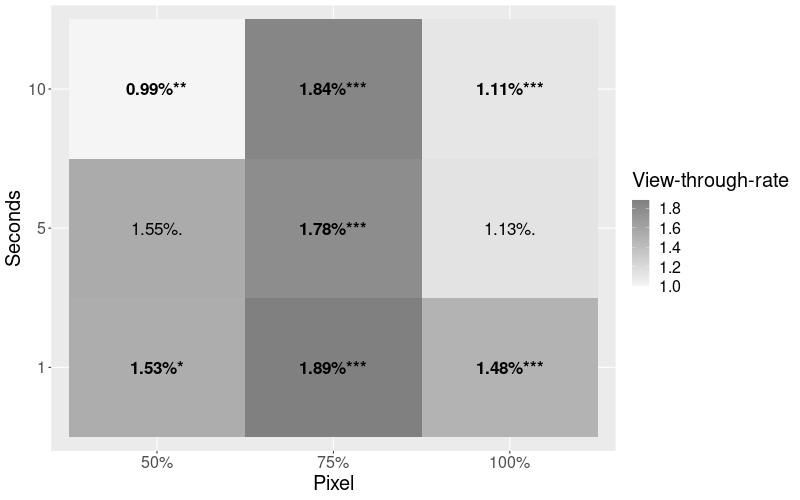}
     \label{fig:obs_heatmap_rawdata}
\begin{minipage}{\textwidth} % choose width suitably
{\footnotesize Notes: 371,952 observations. A view-through is defined as a visit to the advertising website that occurred on the same day as the ad exposure. Share of view-throughs in $<$50/1 condition is 1.3\% (not in graph). Significant differences to $<$50/1 viewability are in bold (given a 5\% significance level based on a chi-square test). Significance codes: $*** < .001$; $** < .01$; $* < .05$; $. < .1$. 
\par}
\end{minipage}
\end{figure}

The matrix in Figure \ref{fig:obs_heatmap_rawdata} shows the view-through rates for nine pixel/seconds combinations, where a darker color indicates a higher view-through rate. Additionally, significant differences to the $<$50/1 condition (which has a view-through rate of 1.3\% and is not depicted in the figure) are in bold, given a 5\% significance level based on a chi-square test. Figure \ref{fig:obs_heatmap_rawdata} reveals that view-through-rates are not necessarily higher for higher viewability rates. More precisely, ads with a pixel percentage-in-view of anywhere between 50\% and 75\% feature the highest view-through rates. In addition, ads with a viewability of 50/10 or 100/10 perform significantly worse than those in the $<$50/1 group in terms of view-through rates, which is at least somewhat unexpected.

\subsubsection*{Empirical Findings}
% latex table generated in R 3.6.1 by xtable 1.8-2 package
% Tue Mar 31 09:02:12 2020

\begin{table}[ht]
\centering
\caption{Influence of Viewability on View-Throughs - Evidence from Observational Data} 
\label{table:observational_results}
\footnotesize
\resizebox{\textwidth}{!}{
\begin{tabular}{lccc}
  \toprule
Variable & Posterior Mean & Standard Deviation & 95\% Credible Interval\\ 
  \midrule
  Intercept & \textbf{-6.97***} & 0.08 & (-7.11, -6.81) \\
  \multicolumn{1}{c}{Viewability} & & &\\
  50\textbackslash 1 & 0.13 & 0.09 & (-0.05, 0.31) \\ 
  50\textbackslash 5 & 0.14 & 0.11 & (-0.08, 0.36) \\ 
  50\textbackslash 10 & \textbf{-0.21***} & 0.08  & (-0.37, -0.05) \\ 
  75\textbackslash 1 & \textbf{0.18***} & 0.05  & (0.08, 0.27) \\ 
  75\textbackslash 5 & 0.13 & 0.08  & (-0.04, 0.28) \\ 
  75\textbackslash 10 & 0.08 & 0.05  & (-0.03, 0.17) \\ 
  100\textbackslash 1 & -0.04 & 0.04 & (-0.12, 0.05) \\ 
  100\textbackslash 5 & \textbf{-0.17***} & 0.06  & (-0.29 -0.05) \\ 
  100\textbackslash 10 & \textbf{-0.15***} & 0.04  & (-0.22, -0.08) \\ 
  \multicolumn{1}{c}{Controls} & & &\\
  Above the Fold & \textbf{0.16***} & 0.03  & (0.10, 0.23) \\ 
  Branding Campaign & \textbf{1.21***} & 0.06  & (1.10, 1.33) \\ 
  Number of Previous Ad Exposures & \textbf{0.88***} & 0.03  & (0.83, 0.93) \\ 
  Number of Previous Ad Exposures squared & \textbf{-0.71***} & 0.03  & (-0.76, -0.65) \\ 
  Area of the Ad & \textbf{1.63***} & 0.07  & (1.49, 1.77) \\
  Area of the Ad squared & \textbf{-2.02***} & 0.07  & (-2.16, -1.87) \\
  Programmatic & \textbf{1.73***} & 0.04  & (1.66, 1.82 \\
  Retargeting & \textbf{2.19***} & 0.13  & (1.93, 2.43) \\
   \bottomrule
\end{tabular}
}

\begin{tablenotes}
      \footnotesize
      \item Notes: 371,952 observations. $<50/1$ Viewability is used as a baseline. We report posterior means of the parameters, along with their 95\% credible intervals based on the posterior quantiles in parentheses. $***$ indicates 99\% credible interval excludes zero. $**$ indicates 95\% credible interval excludes zero. Boldface and $*$ indicate 90\% credible interval excludes zero.
    \end{tablenotes}
\end{table}

In this section, we analyze the effect of viewability on view-throughs in a regression model framework. In the observational study, we analyzed view-throughs (= 1 if a user visits the advertiser's website on the same day as the ad exposure, = 0 otherwise) by specifying a logistic regression model for a binary response variable. The viewability values are represented as dummy variables in the regression equation, while the $<$50/1 group is used as a baseline. We additionally controlled for the ad's placement (above or below the fold), the campaign (branding or performance), the number of previous ad exposures on a user level, as well as the size of the ad. Additionally, we controlled for programmatic and retargeted ads. We employed a Bayesian framework with uninformative priors and Markov chain Monte Carlo techniques. We used four independent chain runs of 500 iterations, with the first 250 iterations discarded as a burn-in sample. The inferences are based on the remaining 250 draws from each chain, which converge well. The coefficient estimates (i.e., posterior means along with their 95\% credible interval) are presented in Table \ref{table:observational_results} and confirm the main statements of the model-free evidence: high viewability values are associated with relatively low view-through rates. Yet, the performance of viewability values with 100\% of their pixels-in-view is even worse than the model-free evidence suggests. Specifically, neither the 100\% viewability values nor any of the 10-second viewability values generate higher predicted view-through rates than the $<50/1$ group.

\begin{figure}[!htbp]
  \caption{Predicted Influence of Viewability on View-through rate - Evidence from Observational Data}
  \centering
    \includegraphics[width=0.8\textwidth]{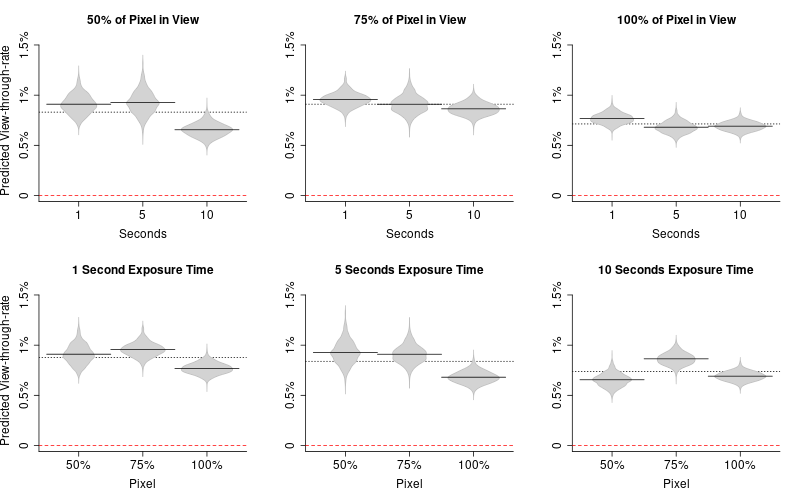}
     \label{fig:observational_beanplots}
\begin{minipage}{\textwidth} % choose width suitably
{\footnotesize Notes: 371,952 observations. Predictions are based on regression results from Table \ref{table:observational_results} and assume above-the-fold placement, branding campaign, no previous ad exposures and a banner size of 300x600 pixel. Solid lines denote respective means, dotted lines denote overall mean.
\par}
\end{minipage}
\end{figure}

Based on our regression results, we built predictions that are presented as bean plots in Figure \ref{fig:observational_beanplots}.\footnote{Please note that view-through rates in Figure \ref{fig:obs_heatmap_rawdata} and \ref{fig:observational_beanplots} differ considerably because predictions are based on users with no previous ad exposures. However, most of the users in the data set exhibit multiple ad exposures.} We chose to arrange the bean plots by showing the posterior densities of the predicted view-through rates in two ways. The first row of plots is organized by pixel percentage-in-view across plots and by exposure duration within plots. The first three plots are thereby intended to demonstrate the impact of exposure duration on predicted view-through rates. The second row is organized by exposure duration across plots and by pixels-in-view within plots; it is intended to show the impact of pixel percentage-in-view on predicted view-trough rates. The first row of plots shows a slightly negative relationship between exposure duration and view-through rate. As pixel percentage-in-view increases the relationship gets weaker. Thus, predicted view-through rates for 100/5 and 100/10 viewability are approximately the same. In contrast to the negative relationship between exposure duration and view-through rates, the second row reveals a slightly different picture. Here we see an inverse u-shaped relationship between pixel percentage-in-view and view-through rates, with the effect expanding as exposure duration increases. Taken together, these findings would encourage advertisers to apply viewability targeting based on the MRC standard, as it is offered by several popular platforms (e.g., Google DoubleClick). Ultimately, ad viewability beyond 75\% or 5 seconds does not seem to perform better. The highest view-through rates come from pixel/second-combinations of 50\%/1, 50\%/5, 75\%/1, and 75\%/5, suggesting that the MRC standard is sufficient for high advertising effectiveness.

\subsection*{Study 2: Online Experiment}\label{sec:study1}
Because ad viewability was not exogenously manipulated in Study 1, we cannot make any causal claims about its effect on advertising effectiveness. Accordingly, we investigated how display ad viewability affects advertising effectiveness in an online experiment in which we controlled for user behavior by exogenously manipulating exposure duration and pixel percentage-in-view. Online participants received a reading comprehension task, with the focal ad displayed on a news website. We decided to use the website of a nationwide news publisher and their most commonly booked ad format since innovative viewability products have first been developed in the news industry (e.g., by the textit{The Financial Times} and \textit{The Economist}). To manipulate viewability exogenously, the ad was sticky (i.e., visibly fixed while the user scrolls), but it either prematurely faded out or was truncated from the right. We randomly assigned the participants to 1 of 10 viewability test cells (see below for details) and asked them to complete a post-exposure survey with various outcome metrics. Thus, we avoided the possibility that user behavior can drive advertising effectiveness and instead measured the causal effect of ad viewability.

\subsubsection*{Experimental Design}
\paragraph{Participants.}\label{sec:study1_participants}
We conducted this online experiment in November 2017, in collaboration with a well-known European ad viewability tracking software provider and a major European news website. To recruit participants, we sent approximately 23,000 invitation emails to the entire student body of a large European business school, offering a chance to win an Apple iPhone 8 in a raffle (the chances of winning the raffle were independent of their experimental responses). We gathered a total of 2,291 participants and predetermined that the data collection would end after 10 days. We had to exclude some participants due to technical problems (e.g., we could not track the viewability of their display ads accurately), leaving a sample of 2,154 respondents. 

\begin{figure}[h!]
  \caption{News Article Website and the Display Ad for Two Experimental Conditions 
  }
  \centering
    \includegraphics[width=\textwidth]{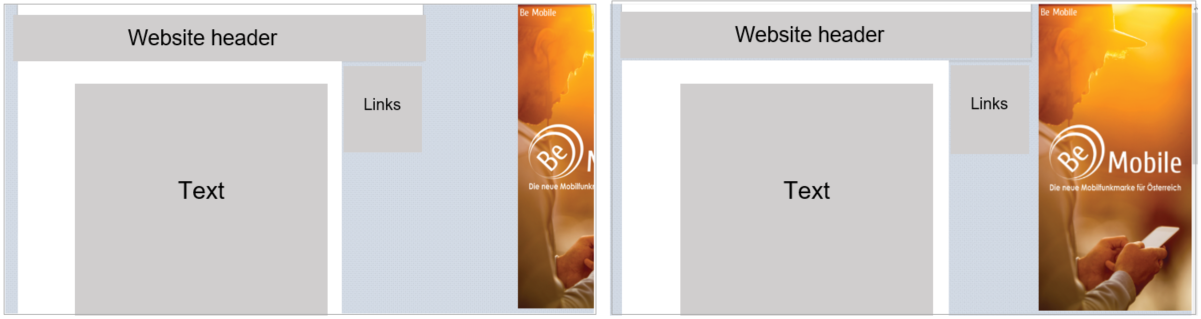}
     \label{fig:website}
\begin{minipage}{\textwidth} % choose width suitably
{\footnotesize Notes: These examples depict conditions with 50\% of the pixels-in-view (left) and 100\% of the pixels-in-view (right). The website identity is masked for anonymity.\par}
\end{minipage}
\end{figure}

\paragraph{Stimuli.}\label{sec:study1_stimuli}
The online display ad represents a half-page ad, with 300 $\times$ 600 pixels (see Figure \ref{fig:website}), which is a standard format \citep{adsense_formats} and frequently appears on the website of our collaborating news publisher. All experimental conditions featured the same ad format and placement (above the fold) for comparability. As we noted previously, the sticky ad remained fixed and visible while users scrolled. Designed exclusively for this study, the ad referred to an unknown mobile telecommunications provider, which did not exist in the market at the time of our study, to avoid the potential confounding effects of using an existing brand. To ensure the applicability of the findings, the brand's name and logo were centrally placed in the ad. Because students represent a key target segment for mobile telecommunications, we thought they would find this brand interesting and relevant. Advertising for and purchases of telecommunication services are also common online, so this distribution channel is well known and accepted. Participants did not see any particular ad more than once.

\paragraph{Randomization into Experimental Conditions.}\label{sec:study1_design}
We randomly assigned participants to 1 of 10 independent experimental conditions, defined by their pixel percentage-in-view and exposure duration in seconds (50/1, 50/5, 50/10, 75/1, 75/5, 75/10, 100/1, 100/5, 100/10, and a full exposure condition).\footnote{Note that, even though the same viewability terms are used in both studies, they describe different viewability values in each of the studies. Whereas 50/1 in the observational study stands for an ad with viewability between 50\% and 74\% of the pixels-in-view and an exposure duration of anywhere between 1 and 4.99 seconds, in the online experiment, it means a manipulation of the ad whereby exactly 50\% of the ad's pixels are in view for one second.} Using a between-subjects design, we studied the effect of ad viewability on advertising effectiveness. Table \ref{table:n_obs_table_study1} details the number of observations and the distribution of selected covariates for each experimental condition. The realized sample size per experimental condition varies between 184 and 246 observations and exceeds current requirements for experimental studies of more than 50 respondents per cell \citep{simmons2014mturk}.

\begin{table}[h!]
\centering
\caption{Number of Observations per Experimental Condition, Study 2}
\label{table:n_obs_table_study1}
\footnotesize
\resizebox{\textwidth}{!}{
\begin{tabular}{p{2.5cm}llllllllllll}
\toprule
& \multicolumn{11}{c}{Experimental condition}\\ 
\midrule
& \textbf{50/1} & \textbf{50/5} & \textbf{50/10} & \textbf{75/1} & \textbf{75/5} & \textbf{75/10} & \textbf{100/1} & \textbf{100/5} & \textbf{100/10} & \textbf{full} & \textbf{overall}  \\ 
\midrule
\textbf{Number of observations}      & 207           & 184           & 223            & 221           & 227           & 188            & 246            & 199            & 237             & 222           & 2154              \\
\textbf{Share of female respondents} & 43\%          & 42\%          & 49\%           & 43\%          & 50\%          & 50\%           & 54\%           & 40\%           & 50\%            & 48\%          & 47\%              \\
\textbf{Mean respondents' age}       & 23y           & 23y           & 23y           & 23y            & 24y            & 23y           & 23y            & 23y            & 23y             & 23y           & 23y               \\
\bottomrule
\end{tabular}
}
\end{table}

We control pixel percentage-in-view and exposure duration in order to test the causal effect of ad viewability on advertising effectiveness. To ensure participants received the treatment as intended, we relied on the technology provided by the ad viewability measurement partner, which can truncate the display ad to the right (such that 50\%, 75\%, or 100\% of the pixels are in view) as well as prematurely fade out the ad (such that the ad remains in view for only 1, 5, or 10 seconds). In one of the experimental conditions, we provided full exposure of the display ad, without any truncation or premature fading out of the display ad. For the truncation to appear as natural as possible, each ad was located on the right side of the page (see Figure \ref{fig:website}). Truncation can occur naturally in a real browsing session when a user does not maximize the browser window (which is often the case on Apple devices) and thereby limits the content of the website (and potential ads) from the right, which we confirmed via informal discussions with industry experts and our viewability measurement partner. Fading out and replacing ads is also becoming increasingly popular with European publishers, who seek to sell single ad slots multiple times. 

Still, some participants' attention may have been drawn to the truncated ads. To rule out potential confounds, we directly asked participants whether they had noticed anything unusual on the experimental treatment page. If they did, we asked them what they considered unusual; among the 2,291 participants, only 3.8\% perceived something “odd” in the ad; across the experimental conditions, the number of participants who noticed anything strange varied between 2 and 15 people. Excluding their data does not influence our results. 

We also asked participants to turn off ad blockers and maximize their browser window. Using the technology provided by our study partner, we are able to identify any users who did not turn off their ad blockers and thus did not receive the ad; these participants were equally distributed across experimental conditions and were excluded from the sample. Furthermore, users were able to adjust their browser window, despite our request that they maximize it. Consequently, some participants did not comply with their randomly assigned, intended viewability value (also called treatment non-compliance). We therefore measured ad viewability as received per experimental group and compared them to ad viewability as intended in all our analyses. Since the results did not differ, we report model-free evidence and regression results for ad viewability as intended. Predictions based on regression results for different data preparation methods are available in the Appendix.

\paragraph{Experimental Procedure.}\label{sec:study1_procedure}
In the email invitation, we asked potential participants to read an article on a major European news website and answer questions about its content. While reading the article, the participants were exposed to a display ad, reflecting one of our experimental treatment groups. The article is identical to an actual article that appeared on the news website; it covers a topic of interest to the target group, includes 390 words, and takes about two minutes to read. To read it completely, participants had to scroll down. They spent an average of 158.8 seconds (2:38 minutes) on the article page. Each experimental session took approximately 10-20 minutes to complete (mean response time = 17.4 minutes). After answering a few questions about the content of the article, participants completed a post-exposure survey. According to common industry practice \citep[e.g.,][]{facebook}, we measured ad recognition on a 3-point scale with a single question: ``Have you seen this ad on the previous news article webpage?'' (1 = no, 2 = not sure, and 3 = yes).

\subsubsection*{Empirical Findings}\label{sec:study1_findings}
\begin{figure}[!htbp]
  \caption{Ad Recognition Shares per Experimental Condition - Evidence from an Online Experiment}
  \centering
    \includegraphics[width=0.8\textwidth]{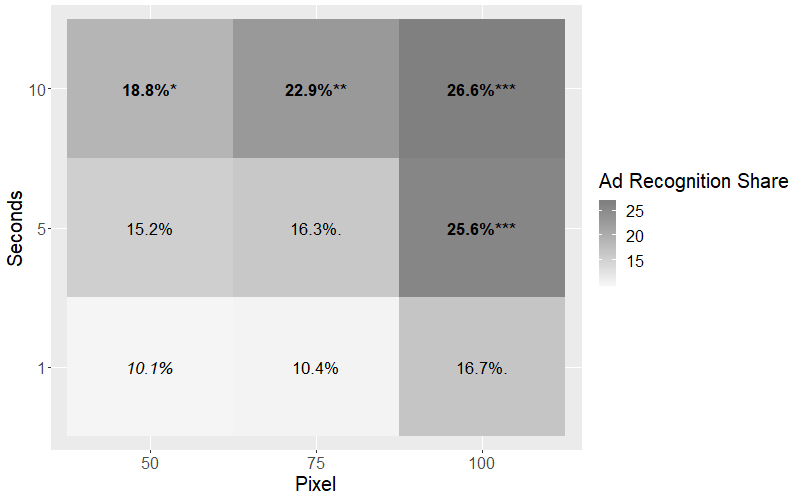}
     \label{fig:online_heatmap_rawdata}
\begin{minipage}{\textwidth} % choose width suitably
{\footnotesize Notes: 2,154 observations. Share of ad recognition in the full exposure condition is 55\% (not in graph). Significant differences to 50/1 viewability are in bold (given a 5\% significance level based on a chi-square test). Significance codes: $*** < .001$; $** < .01$; $* < .05$; $. < .1$.
\par}
\end{minipage}
\end{figure}

Across all experimental conditions with 2,154 participants, 22\% of participants recognized the banner ad in the post exposure survey (`No'=52\%, `I am not sure'=26\%, and `Yes'=22\%). Before we present detailed model results, we offer some model-free evidence according to the randomly assigned experimental conditions. Figure \ref{fig:online_heatmap_rawdata} depicts the share of participants who recognized the experimental banner (i.e., answer = `Yes') per condition. Darker colors indicate a higher ad recognition share while significant differences to a viewability value of 50/1 are depicted in bold (given a 5\% significance level based on a chi-square test). The matrix in Figure \ref{fig:online_heatmap_rawdata} suggests that both increased pixel percentage-in-view and longer exposure durations are associated with a significantly greater share of users who recognize the experimental banner. Thereby, the impact of exposure duration seems to be somewhat stronger (increasing ad recognition shares by 8.7 to 12.5 pp) than the impact of pixel percentage-in-view (increasing ad recognition shares by 6.6 to 10.4 pp). This is also supported by the ad recognition share of the full exposure condition (with a mean exposure duration of 17.7 seconds), which was not included in the figure for better readability. Its ad recognition share of 55\% indicates that increasing exposure duration to more than 10 seconds can have a substantial positive impact on ad recognition. 

The regression results in Table \ref{table:intended_results} largely confirm our results from the model-free evidence. Here, we analyzed viewability and its impact on ad recognition in an ordered logistic regression framework where the answer categories `No', `I am not sure,' and `Yes' are interpreted as a categorical variable with ordered categories. The experimental viewability categories are dummy-coded for the regression and a viewability of 50/1 is used as a baseline. To estimate our model parameters, we again employed Markov chain Monte Carlo sampling in a Bayesian framework with uninformative priors. We ran four independent chains with 500 iterations, where the first 250 iterations were discarded as a burn-in sample and inferences were based on the remaining 250 draws from each chain. The posterior means reported in Table \ref{table:intended_results} indicate that a higher viewability was associated with a higher ad recognition share. Only relatively small viewability values of 50/5, 75/1 and 100/1 did not perform significantly better than the 50/1 baseline condition.

\begin{figure}[!htbp]
  \caption{Predicted Influence of Viewability on Ad Recognition - Evidence from an Online Experiment}
  \centering
    \includegraphics[width=0.8\textwidth]{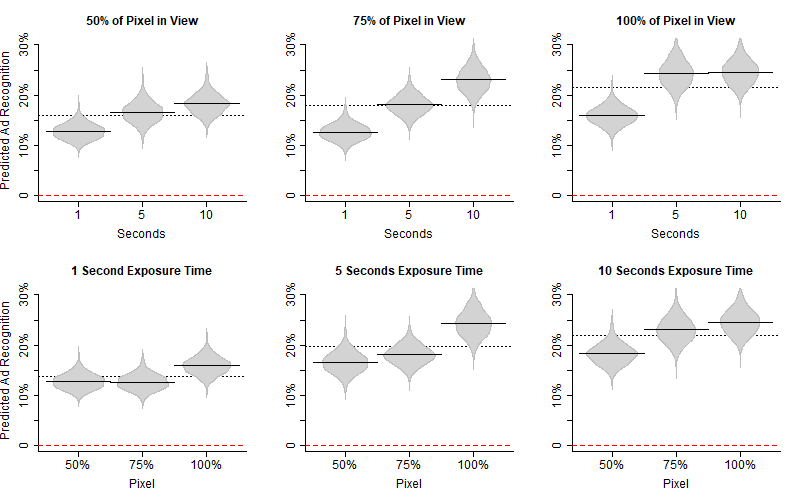}
     \label{fig:intended_beanplots}
\begin{minipage}{\textwidth} % choose width suitably
{\footnotesize Notes: 2,154 observations. Predictions are based on regression results from Table \ref{table:intended_results}. Solid lines denote respective means, dotted lines denote overall mean.
\par}
\end{minipage}
\end{figure}

As in the observational study, Figure \ref{fig:intended_beanplots} illustrates the predictions of ad recognition shares based on posterior densities of our regression results. The corresponding bean plots are arranged such that the first row illustrates the effect of a longer exposure duration and the second row displays the effect of more pixels-in-view. Focusing on the top row first, we find evidence for a clear positive relationship between exposure duration and ad recognition. An increasing exposure duration seems to have the strongest impact on ad recognition with 75\% of the pixels-in-view, as it exhibits the steepest slope. With 100\% of the pixels-in-view, a substantial boost in ad recognition can be attained by increasing exposure duration from 1 to 5 seconds. Further increasing viewability to 100/10, however, does not offer a considerable benefit for advertisers; saturation seems to kick in at this point. Basically, the bottom row of Figure \ref{fig:intended_beanplots} shows a positive effect of increasing pixels-in-view on ad recognition shares. The importance of having the full ad in view, seems to be higher for shorter exposure durations (i.e., 1 and 5 seconds). For an exposure duration of 10 seconds, however, it is enough to have 75\% of the pixels-in-view; a further increase to 100/10 viewability exhibits meager improvements in ad recognition shares. Taken together, we conclude that it is sufficient to have either a long exposure duration or a high pixel percentage-in-view to reach relatively high ad recognition shares.

\begin{table}[ht]
\centering
\caption{Influence of Viewability on Ad Recognition - Evidence from an Online Experiment} 
\label{table:intended_results}
\footnotesize
\resizebox{\textwidth}{!}{
\begin{tabular}{lccc}
  \toprule
Variable & Posterior Mean & Standard Deviation & 95\% Credible Interval\\ 
  \midrule
  \multicolumn{1}{c}{Category Specific Intercepts} & & &\\
  $c_1$ & \textbf{0.64***} & 0.13 & (0.40, 0.91) \\
  $c_2$ & \textbf{1.93***} & 0.14 & (1.66, 2.18) \\
  \multicolumn{1}{c}{Viewability} & & &\\
  50\textbackslash 5 & 0.30 & 0.19 & (-0.08, 0.65) \\ 
  50\textbackslash 10 & \textbf{0.42***} & 0.19  & (0.07, 0.80) \\ 
  75\textbackslash 1 & -0.01 & 0.18  & (-0.37, 0.36) \\ 
  75\textbackslash 5 & \textbf{0.41**} & 0.18  & (0.07, 0.77) \\ 
  75\textbackslash 10 & \textbf{0.72***} & 0.19  & (0.37, 1.08) \\ 
  100\textbackslash 1 & 0.26 & 0.18 & (-0.08, 0.59) \\ 
  100\textbackslash 5 & \textbf{0.79***} & 0.19  & (0.43, 1.15) \\ 
  100\textbackslash 10 & \textbf{0.80***} & 0.18  & (0.47, 1.16) \\ 
  Full & \textbf{2.09***} & 0.18  & (1.74, 2.46) \\ 
   \bottomrule
\end{tabular}
}

\begin{tablenotes}
      \footnotesize
      \item Notes: 2,154 observations. $50/1$  viewability is used as a baseline. We report posterior means of the parameters, along with their 95\% credible intervals based on the posterior quantiles in parentheses.
$***$ indicates 99\% credible interval excludes zero. $**$ indicates 95\% credible interval excludes zero. Boldface and $*$ indicate 90\% credible interval excludes zero.
    \end{tablenotes}
\end{table}

\section*{Discussion and Limitations}\label{sec:summary}
As the online advertising industry grows, advertisers increasingly express their concerns about viewability. Ad viewability refers to an ad's opportunity of being seen online and it is a critical prerequisite of advertising effectiveness. Due to a lack of academic literature on this issue, we offer an initial attempt to rigorously and comprehensively test ad viewability. Through two large-scale empirical studies, we are the first to reveal how ad viewability - comprising both pixel percentage-in-view and exposure duration - affects advertising effectiveness.

Our first study used an observational data set, which is typically available to digital advertisers and publishers to determine campaign effectiveness, with more than 350,000 ad impressions, showing that 30.4\% of the ads have a viewability value below 50/1 and are therefore non-viewable by the MRC standard. A deeper analysis of the observational data suggests that higher viewability values might not always be optimal in generating view-throughs. Specifically, it appears that longer exposure durations ($>$10 seconds) as well as 100\% of the pixels-in-view are inferior to pixel/second-combinations of 50/1, 50/5, 75/1, and 75/5. Such an analysis of the observational data set would therefore encourage advertisers to apply viewability targeting based on the 50/1 standard promoted by the MRC as it appears to be sufficient to achieve high advertising effectiveness. However, this study does not address potential endogeneity concerns, such as user behavior, that might drive ad viewability and correlate with advertising effectiveness.

Consequently, we manipulated ad viewability in a randomized online experiment. Here, we see that higher pixel/second-combinations of 75/10, 100/5, 100/10 show the highest ad recognition rates. Everything below 75\% or 5 seconds seems to perform a lot worse. Thus, we found a discrepancy between observational and experimental data. When accounting for user behavior in a randomized experiment, targeting medium viewability rates, as suggested by our observational study, turns out to be insufficient from an advertising effectiveness stance. The good news for advertisers is though that it does not seem to be necessary to pay for viewability pricing schemes that offer both, a long exposure duration and a high pixel percentage-in-view. Our experiment suggests that either condition is sufficient for reaching relatively high ad recognition shares.

Substantively, our results advance academic literature pertaining to online display advertising and ad exposures. In particular, we purposefully challenge the implicit assumption in academic research that online ads are always viewable. Given that users spend on average about 2-3 seconds on a site, our findings are important for advertisers and publishers alike. First, they enable advertisers to establish target viewability rates more appropriately. Specifically, advertisers will have to understand that the MRC standard is \textit{not} enough from an advertising effectiveness stance. But at least they will not have to maximize both, exposure duration and pixel percentage-in-view, at the same time. Second, our findings encourage publishers to rethink their current viewability products, and offer additional pixel/second thresholds in their dashboards, instead of targeting options based on the 50/1 MRC standard only. Likewise, publishers could adjust their pricing schemes accordingly, charging higher prices for the most effective pixel/second combination. We hope this first step inspires further effort to determine the exact effects of ad viewability on advertising effectiveness. As our research is limited to a very specific setting, further research might test ad viewability in an even broader range of applications. For example, it may be insightful to extend our findings to different ad placements and formats, website types, industries or product categories, as well as to assess ad viewability in a mobile browsing context. Such efforts could further illuminate ad viewability in a broader context of applications and thereby contribute to marketing theory and practice. In addition, one might want to analyze the costs and benefits of targeting certain viewability rates (some might offer lower campaign reach at higher costs). Furthermore, we focused on the effect of viewability on branding metrics -- here, view-throughs and ad recognition -- which both require memory formation. Future research may study the effect of ad viewability on immediate consumer responses, such as clicks. Finally, even though our online experiment already indicates the lower effectiveness of medium ad viewability, future research may want to still explore ad viewability below the 50\%/1-standard.

\clearpage
\bibliographystyle{ormsv080} % outcomment this and next line in Case 1
\bibliography{myrefs} % if more than one, comma separated

\begin{thebibliography}{51}
\expandafter\ifx\csname natexlab\endcsname\relax\def\natexlab#1{#1}\fi
\expandafter\ifx\csname url\endcsname\relax
  \def\url#1{{\tt #1}}\fi
\expandafter\ifx\csname urlprefix\endcsname\relax\def\urlprefix{URL }\fi
\expandafter\ifx\csname urlstyle\endcsname\relax
  \expandafter\ifx\csname doi\endcsname\relax
  \def\doi#1{doi:\discretionary{}{}{}#1}\fi \else
  \expandafter\ifx\csname doi\endcsname\relax
  \def\doi{doi:\discretionary{}{}{}\begingroup \urlstyle{rm}\Url}\fi \fi

\bibitem[{AdSense(2018)}]{adsense_formats}
AdSense, Google. 2018.
\newblock Guide to ad sizes.
\newblock \url{https://support.google.com/adsense/answer/6002621?hl=en}.
\newblock [Last accessed: October 2018].

\bibitem[{AdWords(2018)}]{GoogleAdWords2018}
AdWords, Google. 2018.
\newblock Bid on viewable impressions using viewable cpm.
\newblock \url{https://support.google.com/adwords/answer/3499086?hl=en-GB}.
\newblock [Last accessed: June 2018].

\bibitem[{Albright(2019)}]{albright2019}
Albright, Dann. 2019.
\newblock Benchmarking average session duration: What it means and how to
  improve it.
\newblock
  \url{https://databox.com/average-session-duration-benchmark#difference}.
\newblock [Last accessed: August 2020].

\bibitem[{Baker(2017)}]{Baker2017}
Baker, Jeff. 2017.
\newblock Brafton 2017 content marketing benchmark report.
\newblock
  \url{https://www.brafton.com/blog/strategy/brafton-2017-content-marketing-benchmark-report/}.
\newblock [Last accessed: August 2020].

\bibitem[{Bleier and Eisenbeiss(2015)}]{bleier2015personalized}
Bleier, Alexander, Maik Eisenbeiss. 2015.
\newblock Personalized online advertising effectiveness: The interplay of what,
  when, and where.
\newblock {\it Marketing Science\/} {\bf 34}(5) 669--688.

\bibitem[{Bounie, Morrisson, and Quinn(2017)}]{Bounie2017}
Bounie, David, Val{\'e}rie Morrisson, Martin Quinn. 2017.
\newblock Do you see what i see? ad viewability and the economics of online
  advertising.
\newblock \url{https://papers.ssrn.com/sol3/papers.cfm?abstract_id=2854265/}.
\newblock [Last accessed: September 2018].

\bibitem[{Bounie, Quinn, and Val{\'e}rie(2016)}]{bounie2016advertising}
Bounie, David, Martin Quinn, Morrisson Val{\'e}rie. 2016.
\newblock Advertising viewability in online branding campaigns.
\newblock {\it Available at SSRN 2969891\/} .

\bibitem[{Choi et~al.(2019)Choi, Mela, Balseiro, and Leary}]{choi2019online}
Choi, Hana, Carl~F Mela, Santiago Balseiro, Adam Leary. 2019.
\newblock Online display advertising markets: A literature review and future
  directions.
\newblock {\it Columbia Business School Research Paper\/} {\bf 18}(1).

\bibitem[{Davies(2015)}]{Davies2015}
Davies, Jessica. 2015.
\newblock The economist adopts time-based ad sales.
\newblock \url{https://digiday.com/uk/economist-adopts-time-based-ad-sales/}.
\newblock [Last accessed: June 2018].

\bibitem[{Digishuffle(2019)}]{digishuffle2019}
Digishuffle. 2019.
\newblock 2018 google benchmarks: Bounce rate \& avg. session duration.
\newblock
  \url{https://www.digishuffle.com/blogs/bounce-rate-session-duration-benchmarks/#sessiondurationbenchmarks}.
\newblock [Last accessed: August 2020].

\bibitem[{DoubleClick(2018)}]{doubleclick2018}
DoubleClick. 2018.
\newblock Viewability targeting.
\newblock
  \url{https://support.google.com/bidmanager/answer/6101342?hl=en&ref_topic=2949129}.
\newblock [Last accessed: June 2018].

\bibitem[{Eckles, Gordon, and Johnson(2018)}]{eckles2018field}
Eckles, Dean, Brett~R Gordon, Garrett~A Johnson. 2018.
\newblock Field studies of psychologically targeted ads face threats to
  internal validity.
\newblock {\it Proceedings of the National Academy of Sciences\/}  201805363.

\bibitem[{Elsen, Pieters, and Wedel(2016)}]{elsen2016thin}
Elsen, Millie, Rik Pieters, Michel Wedel. 2016.
\newblock Thin slice impressions: how advertising evaluation depends on
  exposure duration.
\newblock {\it Journal of Marketing Research\/} {\bf 53}(4) 563--579.

\bibitem[{eMarketer(2016)}]{emarketer2016a}
eMarketer. 2016.
\newblock What concerns advertisers about digital media buying?
\newblock
  \url{http://www.emarketer.com/Article/What-Concerns-Advertisers-About-Digital-Media-Buying/1014027}.
\newblock [Last accessed: June 2018].

\bibitem[{Facebook(2020)}]{facebook}
Facebook. 2020.
\newblock Objectives brand awareness.
\newblock
  \urlprefix\url{https://en-gb.facebook.com/business/help/1029827880390718?id=429023050853196}.

\bibitem[{Fang, Singh, and Ahluwalia(2007)}]{fang2007examination}
Fang, Xiang, Surendra Singh, Rohini Ahluwalia. 2007.
\newblock An examination of different explanations for the mere exposure
  effect.
\newblock {\it Journal of consumer research\/} {\bf 34}(1) 97--103.

\bibitem[{Ghose and Todri(2016)}]{ghose2016towards}
Ghose, A, V~Todri. 2016.
\newblock Towards digital attribution: Measuring the impact of display
  advertising on online search behavior, forthcoming.
\newblock {\it MIS Quarterly\/} .

\bibitem[{Goldfarb and Tucker(2011)}]{goldfarb2011online}
Goldfarb, Avi, Catherine Tucker. 2011.
\newblock Online display advertising: Targeting and obtrusiveness.
\newblock {\it Marketing Science\/} {\bf 30}(3) 389--404.

\bibitem[{Goldfarb and Tucker(2014)}]{goldfarb2014standardization}
Goldfarb, Avi, Catherine~E Tucker. 2014.
\newblock Standardization and the effectiveness of online advertising.
\newblock {\it Management Science\/} {\bf 61}(11) 2707--2719.

\bibitem[{Goldstein, McAfee, and Suri(2011)}]{goldstein2011effects}
Goldstein, Daniel~G, R~Preston McAfee, Siddharth Suri. 2011.
\newblock The effects of exposure time on memory of display advertisements.
\newblock {\it Proceedings of the 12th ACM conference on Electronic
  commerce\/}. ACM, 49--58.

\bibitem[{Gordon et~al.(2019)Gordon, Zettelmeyer, Bhargava, and
  Chapsky}]{gordon2019comparison}
Gordon, Brett~R, Florian Zettelmeyer, Neha Bhargava, Dan Chapsky. 2019.
\newblock A comparison of approaches to advertising measurement: Evidence from
  big field experiments at facebook.
\newblock {\it Marketing Science\/} {\bf 38}(2) 193--225.

\bibitem[{Hiemstra(2017)}]{Hiemstra2017}
Hiemstra, Siebe. 2017.
\newblock The history and future of ad viewability tech.
\newblock
  \url{https://www.themarketingtechnologist.co/the-history-and-future-of-display-ad-viewability-tech/}.
\newblock [Last accessed: September 2018].

\bibitem[{Hill et~al.(2015)Hill, Moakler, Hubbard, Tsemekhman, Provost, and
  Tsemekhman}]{hill2015measuring}
Hill, Daniel~N, Robert Moakler, Alan~E Hubbard, Vadim Tsemekhman, Foster
  Provost, Kiril Tsemekhman. 2015.
\newblock Measuring causal impact of online actions via natural experiments:
  Application to display advertising.
\newblock {\it Proceedings of the 21th ACM SIGKDD International Conference on
  Knowledge Discovery and Data Mining\/}. ACM, 1839--1847.

\bibitem[{Houston, Childers, and Heckler(1987)}]{houston1987picture}
Houston, Michael~J, Terry~L Childers, Susan~E Heckler. 1987.
\newblock Picture-word consistency and the elaborative processing of
  advertisements.
\newblock {\it Journal of Marketing Research\/}  359--369.

\bibitem[{IAB(2020)}]{iab2019internet}
IAB. 2020.
\newblock Internet advertising revenue report - full year 2019.

\bibitem[{Johnson, Lewis, and Reiley(2016)}]{johnson2016less}
Johnson, Garrett~A, Randall~A Lewis, David~H Reiley. 2016.
\newblock When less is more: Data and power in advertising experiments.
\newblock {\it Marketing Science\/} {\bf 36}(1) 43--53.

\bibitem[{Joseph(2020)}]{joseph2020}
Joseph, Seb. 2020.
\newblock ‘we must create new proxies’: In the absence of cookies,
  advertisers focus on attention-based metrics.

\bibitem[{Kannan, Reinartz, and Verhoef(2016)}]{kannan2016path}
Kannan, PK, Werner Reinartz, Peter~C Verhoef. 2016.
\newblock The path to purchase and attribution modeling: Introduction to
  special section.

\bibitem[{Lambrecht and Tucker(2013)}]{lambrecht2013does}
Lambrecht, Anja, Catherine Tucker. 2013.
\newblock When does retargeting work? information specificity in online
  advertising.
\newblock {\it Journal of Marketing Research\/} {\bf 50}(5) 561--576.

\bibitem[{Lewis and Rao(2015)}]{lewis2015unfavorable}
Lewis, Randall~A, Justin~M Rao. 2015.
\newblock The unfavorable economics of measuring the returns to advertising.
\newblock {\it The Quarterly Journal of Economics\/} {\bf 130}(4) 1941--1973.

\bibitem[{Manchanda et~al.(2006)Manchanda, Dub{\'e}, Goh, and
  Chintagunta}]{manchanda2006effect}
Manchanda, Puneet, Jean-Pierre Dub{\'e}, Khim~Yong Goh, Pradeep~K Chintagunta.
  2006.
\newblock The effect of banner advertising on internet purchasing.
\newblock {\it Journal of Marketing Research\/} {\bf 43}(1) 98--108.

\bibitem[{Marvin(2014)}]{marvin2014}
Marvin, Ginny. 2014.
\newblock What affects ad viewability? 5 factors from a google study.
\newblock
  \url{https://marketingland.com/affects-ad-viewability-5-factors-google-study-109876}.
\newblock [Last accessed: June 2018].

\bibitem[{Moses(2013)}]{Moses2013}
Moses, Lucia. 2013.
\newblock Ads above the fold are not more viewable.
\newblock
  \url{https://www.adweek.com/brand-marketing/ads-above-fold-are-not-more-viewable-151331/}.
\newblock [Last accessed: June 2018].

\bibitem[{Moses(2015)}]{Moses2015}
Moses, Lucia. 2015.
\newblock The ft has 13 brands on board for time-based ad campaigns.
\newblock
  \url{https://digiday.com/media/financial-times-time-based-ads-impressions/}.
\newblock [Last accessed: June 2018].

\bibitem[{MRC(2014)}]{mrc2014}
MRC. 2014.
\newblock Mrc viewable ad impression measurement guidelines.
\newblock
  \url{http://www.mediaratingcouncil.org/063014%20Viewable%20Ad%20Impression%20Guideline_Final.pdf}.
\newblock [Last accessed: June 2018].

\bibitem[{MRC(2016)}]{mrc2016}
MRC. 2016.
\newblock Mrc mobile viewable ad impression measurement guidelines.
\newblock
  \url{http://measurementnow.net/wp-content/uploads/2015/01/FINAL-062816-Mobile-Viewable-Guidelines-Final-1.pdf}.
\newblock [Last accessed: October 2018].

\bibitem[{Orquin and Wedel(2020)}]{orquin2020contributions}
Orquin, Jacob~Lund, Michel Wedel. 2020.
\newblock Contributions to attention based marketing: Foundations, insights,
  and challenges.

\bibitem[{Pieters and Wedel(2012)}]{pieters2012ad}
Pieters, Rik, Michel Wedel. 2012.
\newblock Ad gist: Ad communication in a single eye fixation.
\newblock {\it Marketing Science\/} {\bf 31}(1) 59--73.

\bibitem[{Rowntree(2016)}]{Rowntree2016}
Rowntree, Lindsay. 2016.
\newblock Current viewability standards don’t equate to effectiveness: Q\&a
  with meetrics \& platform161.
\newblock \url{https://www.exchangewire.com/blog/2016/08/22/44920/}.
\newblock [Last accessed: June 2018].

\bibitem[{Rowntree(2017)}]{Rowntree2017}
Rowntree, Lindsay. 2017.
\newblock 'cost per second' takes viewability to the next level: Q\&a with paul
  kelly, parsec media.
\newblock
  \url{https://www.exchangewire.com/blog/2017/03/07/cost-per-second-takes-viewability-next-level-qa-paul-kelly-parsec-media/}.
\newblock [Last accessed: June 2018].

\bibitem[{Rutz and Bucklin(2012)}]{rutz2012does}
Rutz, Oliver~J, Randolph~E Bucklin. 2012.
\newblock Does banner advertising affect browsing for brands? clickstream
  choice model says yes, for some.
\newblock {\it Quantitative Marketing and Economics\/} {\bf 10}(2) 231--257.

\bibitem[{Sahni(2015)}]{sahni2015effect}
Sahni, Navdeep~S. 2015.
\newblock Effect of temporal spacing between advertising exposures: Evidence
  from online field experiments.
\newblock {\it Quantitative Marketing and Economics\/} {\bf 13}(3) 203--247.

\bibitem[{Sanghvi(2015)}]{Sanghvi2015}
Sanghvi, Nikul. 2015.
\newblock Cost per hour. using a time-based currency for digital advertising.
\newblock
  \url{https://www.slideshare.net/NikulSanghvi/cost-per-hour-using-a-timebased-currency-for-digital-advertising/}.
\newblock [Last accessed: September 2018].

\bibitem[{Simmons(2014)}]{simmons2014mturk}
Simmons, Joe. 2014.
\newblock Mturk vs. the lab: Either way we need big samples.
\newblock {\it Data Colada, datacolada. org/18\/} .

\bibitem[{Vollman(2013)}]{vollman2013}
Vollman, Andrea. 2013.
\newblock Viewability benchmarks show many ads are not in-view but rates vary
  by publisher.
\newblock
  \url{http://www.emarketer.com/Article/What-Concerns-Advertisers-About-Digital-Media-Buying/1014027}.
\newblock [Last accessed: June 2018].

\bibitem[{Vriens(2006)}]{vriens2006handbook}
Vriens, Rajiv Grover~Marco. 2006.
\newblock {\it The handbook of marketing research: uses, misuses, and future
  advances\/}.
\newblock Sage.

\bibitem[{Wang(2017)}]{wang2017viewability}
Wang, Chong. 2017.
\newblock Viewability prediction for display advertising.
\newblock {\it Dissertations\/} .

\bibitem[{Wang et~al.(2015)Wang, Kalra, Borcea, and Chen}]{wang2015viewability}
Wang, Chong, Achir Kalra, Cristian Borcea, Yi~Chen. 2015.
\newblock Viewability prediction for online display ads.
\newblock {\it Proceedings of the 24th ACM International on Conference on
  Information and Knowledge Management\/}. ACM, 413--422.

\bibitem[{Wang et~al.(2017)Wang, Kalra, Zhou, Borcea, and
  Chen}]{wang2017probabilistic}
Wang, Chong, Achir Kalra, Li~Zhou, Cristian Borcea, Yi~Chen. 2017.
\newblock Probabilistic models for ad viewability prediction on the web.
\newblock {\it IEEE Transactions on Knowledge and Data Engineering\/} {\bf
  29}(9) 2012--2025.

\bibitem[{Wang et~al.(2018)Wang, Zhao, Kalra, Borcea, and
  Chen}]{wang2018webpage}
Wang, Chong, Shuai Zhao, Achir Kalra, Cristian Borcea, Yi~Chen. 2018.
\newblock Webpage depth viewability prediction using deep sequential neural
  networks.
\newblock {\it IEEE Transactions on Knowledge and Data Engineering\/} {\bf
  31}(3) 601--614.

\bibitem[{Zhang et~al.(2015)Zhang, Pan, Zhou, and Wang}]{zhang2015empirical}
Zhang, Weinan, Ye~Pan, Tianxiong Zhou, Jun Wang. 2015.
\newblock An empirical study on display ad impression viewability measurements.
\newblock {\it arXiv preprint arXiv:1505.05788\/} .

\end{thebibliography}

\clearpage
\begin{appendix}
\section*{Appendix}
\setcounter{figure}{0} \renewcommand{\thefigure}{A\arabic{figure}} 

\section*{Measuring Display Ad Viewability}\label{sec:measuringviewability}
Consider a scenario: A user visits a news website and an ad gets served, which counts as an ad impression in existing business models. Actual display ad viewability can be measured with two main approaches (for additional measurement methods, see \citealt{Hiemstra2017}). 
First, geometric triangulation uses information about the position of the corners of the browser viewport (i.e., area of the page currently shown on screen) relative to the page, as well as the position of the corners of the ad slot relative to the page. A JavaScript that loads alongside banner ads uses this information to determine if and to what extent a display ad is being shown. As soon as the ad reaches 50\% of its size, the JavaScript starts tracking the time until the banner ad falls below this threshold. If the 50\% banner ad remains within the viewport for 1 continuous second, it is considered viewable according to the MRC. 
Second, the browser optimization method leverages the internal optimization processes that browsers use to reduce page loading time. These optimization processes allocate more resources to loading elements that appear immediately within the viewport or its proximity. As the banner ad gets closer to the viewport, it gets loaded in by the browser at a faster rate. Browser optimization keeps track of these internal processes to determine the position of the banner ad relative to the viewport. The measure of the on-screen time is similar to that used by geometric triangulation.

Geometric triangulation offers better accuracy than browser optimization, which might reflect the influence of slow Internet connections or computers on page loading speed. However, geometric triangulation is not applicable when the ad slot is embedded in an iFrame, which serves the display ad on a different page with a different URL and thereby prevents the JavaScript from obtaining information about the position of the display ad relative to the viewport. Therefore, viewability measurement providers rely on both methods to determine display ad viewability.\footnote{Desktop and mobile browser ad viewability both rely on geometric triangulation and browser optimization, yet measuring mobile app viewability is generally more difficult. We focus on display ad viewability, measured for a desktop computer (browser).}

\section*{Robustness Check}\label{robust}
\begin{figure}[!htbp]
  \caption{Comparison of Predicted Ad Recognition for 3 Different Data Preparation Methods to Investigate Treatment Non-Compliance}
  \centering
    \includegraphics[width=0.8\textwidth]{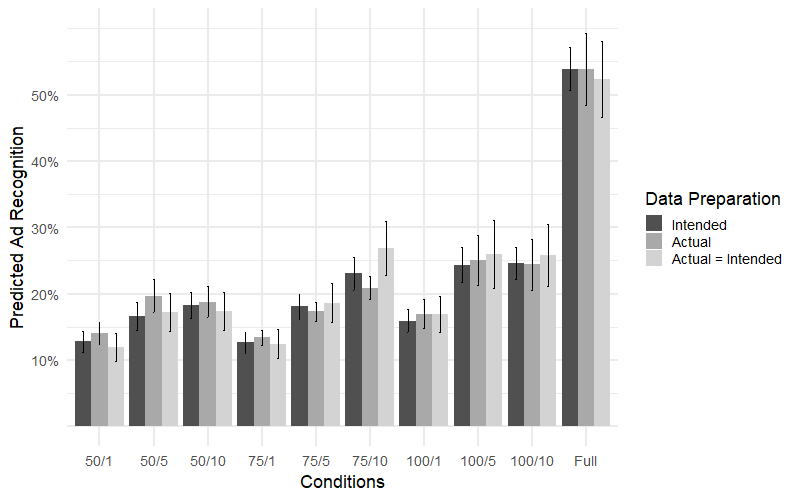}
     \label{fig:robustnesscheck}
\begin{minipage}{\textwidth} % choose width suitably
{\footnotesize Notes: Predictions are based on posterior mean regression results. One standard deviation error bars are reported. `Intended' corresponds to ex-ante dummy coding of viewability values, i.e., category dummies are coded as originally intended in the random assignment of the online experiment (n = 2,154). `Actual' corresponds to ex-post dummy coding of viewability values, i.e., irrespective  of the intended assignment of the experiment but only based on the actual viewability values of the ad (n = 2,154). `Actual = Intended' corresponds to a data set which was filtered for observations where actual viewability equals intended viewability (n = 921).
\par}
\end{minipage}
\end{figure}

The results of the online experiment we presented in the main text were based on experimental conditions to which participants were randomly assigned, irrespective of the actual viewability of the display ad. However, as participants did not always strictly follow the instructions to maximize their browser window, ad viewability as received can differ from ad viewability as intended in some cases (also called treatment non-compliance). Therefore, we present predictions based on three different data preparation methods, i.e., (1) `Intended', (2) `Actual,' and (3) `Actual = Intended', as a robustness check. `Intended' thereby corresponds to ex-ante dummy coding of the viewability values, which means category dummies for the regression are coded as originally assigned in the experiment. In the `Actual' data preparation method, dummies were ex-post coded, i.e., corresponding to viewability values that participants actually reached, irrespective of the original random assignment in the experiment. In the third data preparation method `Actual = Intended', viewability values were strictly filtered for observations that actually fulfilled the intended condition. Figure \ref{fig:robustnesscheck} shows predictions based on regression outcomes for the three different data preparation methods along with their one standard deviation error bars. The figure shows that the main findings of our study are not affected by the choice of the data preparation method as the predicted ad recognition shares are roughly the same for each viewability value, irrespective of the chosen approach.

\end{appendix}

\end{document}